\documentclass[aps,prl,twocolumn,a4paper,10pt, longbibliography]{revtex4-1}

\usepackage[T1]{fontenc}		
\usepackage[english]{babel}		
\usepackage[latin1]{inputenc}	
\usepackage{times}

\usepackage{amsmath}			
\usepackage{amssymb}			
\usepackage{bbm}				
\usepackage{mathtools}
\usepackage{gensymb}
\usepackage{simpler-wick}
\usetikzlibrary{tikzmark}
\DeclarePairedDelimiterX\Dirbraket[3]{\langle}{\rangle}%
{#1\,\delimsize\vert\,\mathopen{}#2\,\delimsize\vert\,\mathopen{}#3}

\usepackage[colorlinks=true, a4paper=true, pdfstartview=FitV,linkcolor=blue, citecolor=blue, urlcolor=blue]{hyperref}

\usepackage{graphicx}			
\usepackage{graphics}			
\DeclareGraphicsExtensions{.pdf}
\usepackage{color}		

\newcommand{\bea}{\begin{eqnarray}}
\newcommand{\eea}{\end{eqnarray}}

\newcommand{\bk}{\mathbf{k}}

\begin{document}

\title{Disconnected and multiply connected spectra in the 2D attractive Hubbard model}

\author{Johan~Carlstr\"om }
\affiliation{Department of Physics, Stockholm University, 106 91 Stockholm, Sweden}
\date{\today}

\begin{abstract}
Fermi gases and liquids display an excitation spectrum that is simply connected, ensuring closed Fermi surfaces. In strongly correlated systems like the cuprate superconductors, the existence of open sheets of Fermi surface known as Fermi arcs indicate a distinctly different topology of the spectrum with no equivalent in Fermi liquid theory. 
Here, we demonstrate a generic mechanism by which correlation effects in fermionic systems can change the topology of the spectrum. 
Using diagrammatic Monte Carlo simulations, we demonstrate the existence of disconnected and multiply connected excitation spectra in the attractive Hubbard model in the BCS-BEC cross-over regime. These topologically nontrivial spectra are a prerequisite for Fermi arcs.  
\end{abstract}
\maketitle

Landaus Fermi liquid theory \cite{landau1980course} is the standard model through which we understand interacting electrons in normal metals. 
In this paradigm, electronic states evolve adiabatically with increasing interactions so that there remains a direct correspondence between the states in a non-interacting Fermi gas, and the quasi-particles of the interacting system. 
A key consequence of this relationship is that the excitation spectrum of the interacting system inherits the topology of the bands associated with the noninteracting state. 
In the absence of gap-closing points, the energy bands of Fermi gases are generally simply connected sets, and so are consequently the spectra of Fermi liquids. 
This, in turn, implies a Fermi surface that is closed (this point also holds with nodes in the spectrum). 
Strongly correlated systems often display phenomena that fall decidedly outside of the Fermi liquid regime. In the cuprates, superconductivity is nucleated from a pseudogap state with open sheets of Fermi surface, which persist over  a wide range of doping levels \cite{doi:10.1126/science.aaw8850}.  
The physical origin of these Fermi arcs remains highly contested.

It has been observed in the cuprates that superconducting fluctuations persist above the critical temperature \cite{Kondo2015, PhysRevX.11.031068, Bergeal2008}, and it has been proposed that this fact may explain the origin of the pseudogap state \cite{Seo2019}.
This in turn raises key questions about the pairing regime, which also remains disputed: If the cuprates are BCS-like, then the fluctuating region should be understood in terms of a paired state without global phase coherence \cite{https://doi.org/10.48550/arxiv.2210.13478}. 
In the BEC limit, the electrons form bound pairs which give rise to a bosonic normal liquid at temperatures far above $T_c$ \cite{PhysRevB.99.104507}. The onset of superconductivity would then occur as these pairs condense at a much lower temperature. While these two scenarios are often both referred to by the term ``preformed pairs'', they are remarkably different. Between these two extrema lies the an extensive BCS-BEC crossover regime \cite{PhysRevLett.129.017001}.

A directly opposing point of view is that preformed pairs have no part in the emergence of Fermi arcs, and that the pseudogap and paired states are instead antagonistic to each other. 
ARPES imaging is claimed to show direct competition between superconductivity, and a distinctly different order parameter that is associated with the pseudogap state \cite{Hashimoto_2010,Hashimoto2015}. 
A candidate for this order parameter is provided by a breaking of translation symmetry \cite{PhysRevLett.101.207002}, which is observed in STM imaging \cite{Wise_2008,Hoffman2002}.

Theoretically predicting the existence of Fermi arcs in model Hamiltonians is challenging due to a lack of reliable numerical techniques for strongly correlated fermions. 
Nonetheless, recent variational Monte Carlo calculations suggest that the pseudogap physics observed in the cuprates is at least qualitatively captured by the single band Hubbard model. For Hubbard clusters up to $64$ sites, Fermi arcs are observed at a carrier concentration of $6.25\%$, and remnants of these are present at $12.5\%$ doping \cite{https://doi.org/10.48550/arxiv.2209.08092}.
This may be compared to the cuprates, where pseudogap physics persist up to a carrier concentration of  $\sim 20\%$ \cite{doi:10.1126/science.aaw8850,Badoux2016}. 
The existence of Fermi arcs in a simple model Hamiltonian like the Hubbard model is encouraging since it may indicate that this is a generic phenomena. 

\begin{figure}[!htb]
\includegraphics[width=\linewidth]{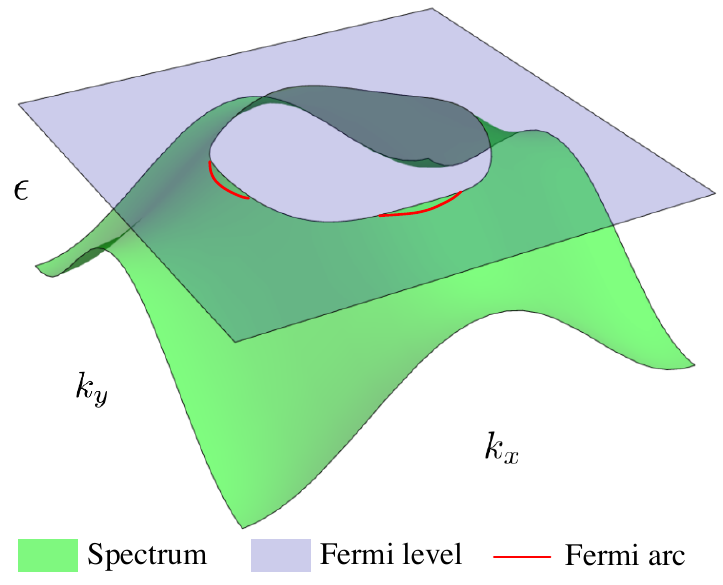}
\caption{
{\bf Relationship between spectral topology and Fermi arcs.} 
The multiply connected spectrum intersects the Fermi level on a set of open and disconnected lines which constitute Fermi arcs. By contrast, a simply connected spectrum, must necessarily intersect the Fermi level on a set of closed lines. This implies that a topologically nontrivial spectrum is a prerequisite of Fermi arcs. 
}
\label{illustration}
\end{figure}

A second theoretical challenge is to qualitatively explain how Fermi liquid theory fails in strongly correlated systems, and connect this insight with the emergence of Fermi arcs. Here, a key observation is that a simply connected excitation spectrum does not permit 
 open sheets of Fermi surface. This relationship implies that the electronic state's adiabatic dependence on interaction strength must necessarily break down in such a way that the connectivity of the spectrum changes, see also Fig. \ref{illustration}.

In this work, we discuss how strong interactions can give rise to non-Fermi-liquid phases which are characterized by band fractionalization \cite{doi:10.7566/JPSJ.88.024701}. Using the attractive-interaction Hubbard model as an example, we demonstrate that that the operators associated with these fractional bands exhibit vanishing phase spaces in parts of the Brillouin zone, which leads to disconnected or multiply connected excitation spectra.  
These topologically nontrivial spectra are a fundamental prerequisite for the existence of Fermi arcs.

{\it Band fractionalization and spectral topology}---To illustrate the breakdown of Fermi liquid theory, we consider the attractive Hubbard model (AHM), which is given by
\bea
H=\sum_{\langle ij\rangle \sigma}t c^\dagger_{i\sigma}c_{j\sigma}+\sum_i (U n_{i\downarrow}n_{i\uparrow}-\mu n_i),\; U<0.\label{Hubbard}
\eea
Because of the interaction, the energy bands are generally split into two sub-bands, \cite{PhysRevB.18.3453}, a phenomena that is also referred to as band fractionalization \cite{doi:10.7566/JPSJ.88.024701}. For strong contact interaction, these sub-bands are generally singlon-like and doublon-like respectively, prompting us to introduce the corresponding operators and associated spinors:
\bea\nonumber
c_{i\sigma}^\dagger=s_{i\sigma}^\dagger+d_{i\sigma}^\dagger,\;
 s_{i\sigma}^\dagger=c_{i\sigma}^\dagger(1-n_{i\bar{\sigma}}), \;d_i^\dagger=  c_{i\sigma}^\dagger  n_{i\bar{\sigma}}\\
\Psi_{i\sigma}^\dagger = \big[s_{i\sigma}^\dagger \;\; d_{i\sigma}^\dagger \big],\;\;\;
\Psi_{i\sigma}=
\begin{bmatrix}
     s_{i\sigma}      \\
    d_{i\sigma}     \label{QP}  
\end{bmatrix}.
\eea
Here, $s^\dagger$ and $d^\dagger$ are the singlon and doublon creation operators while $\bar{\sigma}=-\sigma$. 
We can then define a ``quasi-particle'' (QP) greens function based on the outer product of the spinors:
\bea
\Gamma_\sigma(x_2-x_1)=\langle T_\tau\Psi_{i\sigma}^\dagger(x_1)\otimes \Psi_{i\sigma}(x_2)\rangle,
\eea
from which the ordinary electronic Greens function is obtained by the summation
\bea
G_\sigma(x)=\sum_{\alpha\beta}\Gamma_{\alpha\beta\sigma}(x).
\eea
In the atomic limit, the QP Greens function is diagonal, with a frequency space representation given by
\begin{equation}\label{QPG}
\Gamma_\sigma^A(\omega)=
\begin{bmatrix}
     \frac{1+e^{\mu}}{Z_A}\frac{1}{i\omega+\mu}  &  0      \\
    0  &  \frac{e^{\mu}+e^{2\mu-U}}{Z_A}\frac{1}{i\omega+\mu-U}       
\end{bmatrix}.
\end{equation}
Here, the energy is for simplicity given in units of the temperature (corresponding to the case of unit temperature).
The Greens function (\ref{QPG}) resembles that of a two-component system, except that it is rescaled by two  ``quasiparticle weights''. 
To pursue this analogy we introduce the weight $W$ according to
\begin{equation}\label{Wform}
W=
\begin{bmatrix}
     \frac{1+e^{\mu}}{Z_A}  &  0      \\
    0  &  \frac{e^{\mu}+e^{2\mu-U}}{Z_A}       
\end{bmatrix}
= w_0\sigma_0+w_z\sigma_z, 
\end{equation}
where we note that (\ref{Wform}) must satisfy
\bea
w_0\ge |w_z| \label{Wreq}.
\eea
In the limit $ w_z\to w_0$, the system is effectively Gutzwiller projected, and doublons can be regarded as ``forbidden''. In this scenario, the doublon operators can be said to have a vanishing phase space in the sense that they have a domain or codomain which does not overlap with the sub-space on which we project. The same can be said abut the singlon operator in the limit $w_z\to- w_0$. In these cases, the doublon or singlon parts do not contribute to the Greens function, and thus not to the spectrum either. 

We may then express the atomic Greens function (\ref{QPG}) in terms of a reweighted two-component system according to
\bea
\Gamma_\sigma^A(\omega) =\frac{W}{i\omega-V}, \;\;\; V=\Big[\frac{U}{2}-\mu\Big]\sigma_0-\frac{U}{2}\sigma_z,
\eea
where $V$ is the effective two-component Hamiltonian. 

Next, we
note that the tunneling term may be written
\bea
tc_{i\sigma}^\dagger c_{j\sigma}=\Psi^\dagger_{i\sigma}K\Psi_{j\sigma},\; K=t(\sigma_0+\sigma_x).
\eea
Thus, including the first correction of the strong-coupling expansion \cite{PhysRevB.103.195147}, we 
obtain a Greens function 
\bea\nonumber
\Gamma_\sigma(\omega)=\Gamma_\sigma^A(\omega)+\Gamma_\sigma^A(\omega) K(\bk)\Gamma_\sigma^A(\omega)+...\\
=\frac{1}{i\omega-V-WK(\bk)}W.\label{Dyson1}
\eea
At this point, the effective two-component Hamiltonian $H_e=V+WK(\bk)$ is no longer diagonal, and the dispersion thus mixes the singlon and doublon components. Additionally, $H_e$ is non-Hermitian, and does not generally exhibit an orthonormal eigenbasis. However, due to a combination of $\mathcal{P}\mathcal{T}$ symmetry and the condition (\ref{Wreq}), the eigenvalues remain real. 

Due to the factor $W$, the spectral weight of the two sub-bands are generally not equal, and one of them may even vanish asymptotically. 
This points is central to the spectral topology: If we conduct a strong coupling expansion to higher order, then we will find that the QP weight $W$ becomes momentum dependent. If the phase space for a sub-band operator of the type (\ref{QP}) vanishes in part of Brillouin zone, then so does the corresponding spectral weight, implying that the spectrum is no longer simply connected. 
Strong-coupling expansion by hand is however not feasible beyond first order, and to explore this concept we have to employ numerical techniques.  

{\it Numerical treatment}---To test the preceding conjecture, we employ bold-line diagrammatic Monte Carlo simulations, specifically focusing on the attractive Hubbard model in the BCS-BEC cross over regime. This method is based on stochastic sampling of Feynman type graphs \cite{Van_Houcke_2010}, and is unbiased in the sense that the only systematic source of error is truncation of the series. For a convergent series, asymptotically exact results are obtained directly in the macroscopic limit. 
To be able to address systems with strong interactions we use a particular formulation known as strong-coupling diagrammatic Monte Carlo (SCDMC) \cite{PhysRevB.103.195147,0953-8984-29-38-385602,PhysRevB.97.075119,carlstrom2021spectral,PhysRevResearch.4.043126}, where the diagrammatic elements are connected vertices of propagating electrons that are non-perturbative in $U$. The computational protocol employed here is outlined in detail in \cite{PhysRevB.103.195147}.

 \begin{figure*}[!htb]
 \hbox to \linewidth{ \hss
\includegraphics[width=\linewidth]{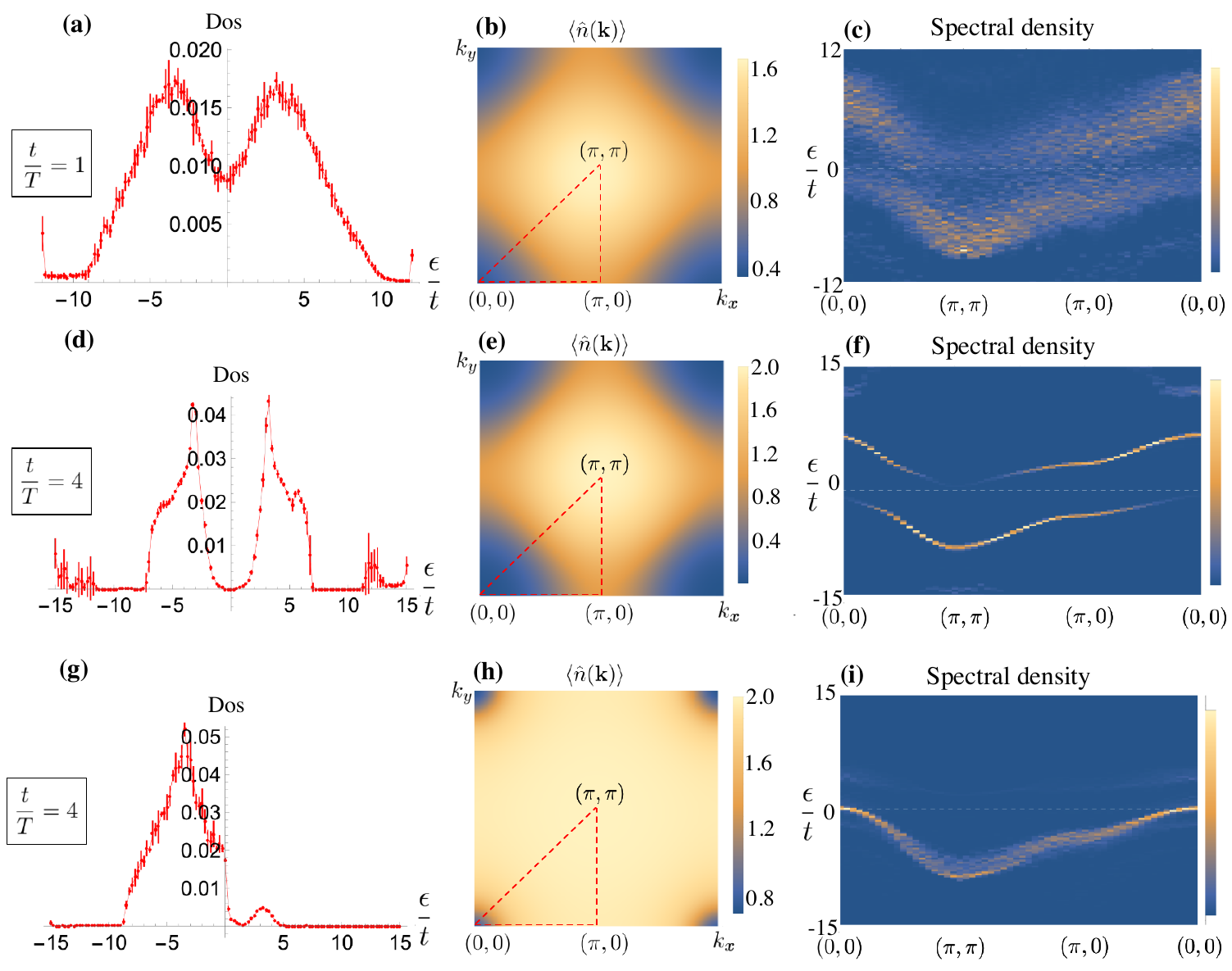}
 \hss}
\caption{
{\bf Spectra and equation of state} for the attractive Hubbard model with $U=-5|t|$, at temperatures of $t/T=1$ (a-c) and $t/T=4$ (d-i). The figures (a-f) corresponds to half-filling, while (g-i) corresponds to $\langle \hat{n}\rangle\approx 1.88$. At high temperature, the spectrum (a) reveals a suppression of the density of states at the Fermi level. The particle density (b) exhibits a minimum at $\bk=(0,0)$ with $\langle \hat{n}\rangle\approx 0.4$ and a maximum at $\bk=(\pi,\pi)$ with $\langle \hat{n}\rangle\approx 1.6$. The momentum-resolved spectral density (c) taken along the dashed line in (b), reveals two sub-bands. 
Decreasing the temperature, the density of states (d) vanishes at the Fermi level, implying that the system is gapped with respect to fermionic excitations. The particle density (e) now has minima and maxima close $0$ and $2.0$ respectively. The spectral density (f) reveals sharp families of excitations with a spectral weight that is strongly dependent on momentum and almost vanishes in part of the Brillouin zone.
Increasing the particle density to $\langle \hat{n}\rangle\approx 1.88$, the density of states (g) reveals a large peak that is doublon-like, and a much suppressed peak corresponding to singlons. The peaks are well separated, and the density of states vanishes at $\epsilon\approx 1.5t$. 
The spectral density reveals a large doublon-like peak, though the singlon peak has a presence mainly near $\bk=(0,0)$.
This data was obtained using an expansion order $O=6$. 
}
\label{spectra}
\end{figure*}

In SCDMC, the expansion parameter is the hopping integral $t$. The principal observable that we compute is the polarization operator of the hopping integral, here denoted $\Pi_t(\omega,\bk)$. From the polarization operator, we obtain the dressed hopping integral via the Bethe Salpiter equation:

\bea
\tilde{t}(\omega,\bk)=\frac{1}{t^{-1}(\bk)-\Pi_t(\omega,\bk)}.\label{bethe}
\eea 
We expand in the dressed hopping $\tilde{t}$, while retaining only the skeleton diagrams. By iterating until convergence, we obtain a self-consistent solution for $\tilde{t}$ which implicitly takes into account certain classes of diagrams to infinite order. 

The Greens function of the interacting system is closely related to the dressed hopping integral, and can be obtained from the equation
\bea
G(\omega,\bk)=\frac{1}{\Pi_t^{-1}(\omega,\bk)-t_\bk}. \label{boldG}
\eea
To the lowest order, the polarization operator is given by the atomic-limit Greens function, meaning that eq. (\ref{Dyson1}) is reproduced.
We conduct a self-consistent summation of all diagrams to order $7$ which permits us to asses convergence properties of the series--for a discussion, see Appendix I.

We compute a discrete approximation for the spectrum using numerical analytical continuation
\cite{PhysRevB.95.014102}: First, we define a spectral reconstruction of the Greens function and a corresponding error metric according to 
\bea\label{GR}
G_R(\tau,\bk)&=&\sum_{n=1}^{n_{\text{max}}} A_n(\bk)\frac{e^{-\epsilon_n \tau}}{1+e^{\beta\epsilon_n}},\;\;\; \tau<0,  \\
\Delta[\bk,\{A_n(\bk)\}] &=&\sqrt{\frac{1}{\beta}\int d\tau [G(\tau,\bk)-G_R(\tau,\bk)]^2}.\;\;\;\;\label{Delta}
\eea
We use $n_{\text{max}}=121$ as a compromise between accuracy and computational cost. 
To obtain the best estimate for the spectral function $A(\bk)$, we minimize the error metric $\Delta$ through a process of simulated annealing followed by a line-search tecnhique:
In the first stage, we use Monte Carlo to update $\{A_n(\bk)\}$ with an acceptance ration $\sim e^{-\kappa \Delta}$, while successively increasing the inverse pseudo temperature $\kappa$. In the second stage, we minimize $\Delta$ using Newton-Raphson. This reduces the error only very slightly, but tends to result in a smoother spectrum. 

From the spectrum, we obtain a (discretized) estimate for the density of states via the integral
\bea
\text{dos}(\epsilon_n)=\int \frac{d\bk}{(2\pi)^D}A_n(\bk).\label{dos}
\eea
The normalization of Eq. (\ref{GR}) is such that the summations over $A_n$ and dos$(\epsilon_n)$ are unity. 

We consider the Hubbard model with an attractive contact interaction given by $U=-5|t|$, at temperatures $t/T=1$ and $t/T=4$. We examine the cases of half-filling and a particle density of $\langle \hat{n}\rangle\approx 1.88$. The results of our simulations are summarized in Fig. \ref{spectra}.

 At half-filling and a higher temperature of $t/T=1$, we find that the density of states (a) has a minimum at the Fermi level, though the system remains gapless. The momentum-resolved particle density (b) attains minima and maxima at $\sim 0.4$ and $\sim 1.6$. The spectral density (c) exhibits two smeared sub-bands, with densities that are visibly momentum-dependent. 
 
 Reducing the temperature, the density of states (d) vanishes at the Fermi level, indicating that the system is gapped against fermionic excitations. The particle density extrema (e) are now close to $0$ and $2.0$ respectively. The spectral density (f) is sharply peaked, with a weight that is strongly dependent on momentum. 
 
 If we also increase the particle density, then the upper sub-band is strongly suppressed as a result (g). The system is now completely filled in a large fraction of the Brillouin zone (h), and the lower sub-band carries most of the spectral weight (i).

 The momentum-dependent spectral weights can be understood from the fact that the two sub-bands originate in singlon-like and doublon-like degrees of freedom:  
 For sufficiently strong attraction, the Hubbard model prefers to have occupation numbers of $0$ or $2$.  Singly occupied sites are situated at high energy, implying that the upper sub-band is singlon-like. At small momenta, $\bk\approx (0,0)$, the particle density is smaller, and the singlon operator has a substantial phase space allowing for a high spectral density. Near $\bk=(\pi,\pi)$, the particle density approaches $2$, meaning that the phase space for the singlon operator vanishes, along with the spectral weight of this sub-band. For the doublon-like component, the situation is the opposite, with a vanishing spectral density when the density is small. 
 
To quantify the suppression of the spectral density, we define the spectral weight of a sub-band according to
\bea
\rho(\bk) = \sum_{n=n_{\text{min}}}^{n=n_{\text{max}}}  A_n(\bk),\label{specw}
\eea
where the range of indices $n$ should be taken to include the entire sub-band, but nothing else. 
At a temperature of $t/T=4$ and halffilling, the system remains gapped so that we can identify the upper sub-band with positive energies and the lower sub-band with negative energies.
Doping the system, the two sub-bands are still well separated with the density of states vanishing at $\epsilon\approx 1.5 t$, suggesting we use this energy as the dividing point. At the higher temperature, the two sub-bands overlap. We can still calculate spectral weights based on $\epsilon=0$ as our dividing point, though this approximation may slightly underestimate the spectral weight at its minimum, while overestimating it at the maximum.

 \begin{figure}[!htb]
\includegraphics[width=\linewidth]{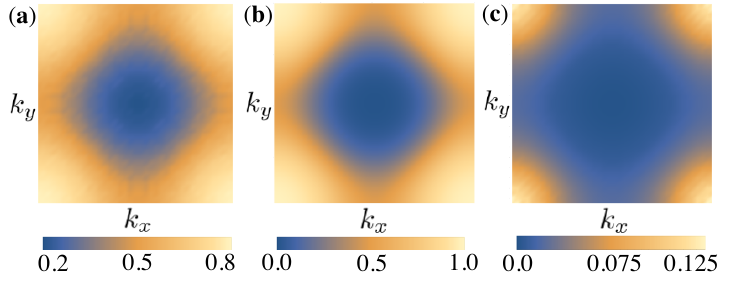}
\caption{
{\bf Spectral weight} of the singlon-like sub-band, obtained from equation (\ref{specw}). 
At half-filling and a temperature of $t/T=1$ (a), the weight is suppressed near $\bk=(\pi,\pi)$ and reaches a minimum of $\approx 16\%$. 
Reducing the temperature (b), this minimum falls below $1\%$.
 Increasing the particle density to $\langle \hat{n}\rangle\approx 1.88$ (c), the spectrum retains a finite weight near $\bk=(0,0)$ but almost vanishes elsewhere. 
 The strong suppression of the spectral weight at certain momenta can be understood from a vanishing phase space of singlon-like excitations.   
}
\label{weight}
\end{figure}

 The spectral weight of the singlon-like component is shown in Fig. \ref{weight}. At a temperature of $t/T=1$ and half-filling (a), the singlon-like component is suppressed to $\approx 16\%$ at $\bk\approx (\pi,\pi)$. At a temperature of $t/T=4$ (b), this minimum drops below $1\%$.
The strong temperature dependence is consistent with the notion of a vanishing phase space for the singlon operator: At $\bk=(\pi,\pi)$, the system has a preference for double occupation, and the singlon operator can only act in the presence of thermal fluctuations. As the temperature is reduced, these are exponentially suppressed together with the spectral weight. Asymptotically, this results in a multiply connected spectrum which lacks states in part of the Brillouin zone. 
 Increasing the particle density (c), the spectral weight attains a maximum at $\bk=(0,0)$ while asymptotically vanishing between these. The result is a disconnected spectrum. 

It should be noted that we do not reach the point where the spectrum completely vanishes since we are limited to finite temperatures. Diagrammatic Monte Carlo generally requires that the series converges, and this is often not the case at sufficiently low temperatures. Real condensed matter systems are also generally realized at finite temperature. However, thermal fluctuations are exponentially suppressed with the inverse temperature. If the relevant energy scale is large compared to the temperature, then we can for all practical purposes regard the systems as being in the asymptotic limit where the spectral density vanishes in part of the Brillouin zone. 
 Once the spectrum has a nontrivial connectivity, there are no topological obstacles to an intersection with the Fermi level that is an open line in 2D, as shown in Fig. \ref{illustration}, or an open surface in 3D. 

{\it Conclusions}---In non-Fermi-liquids, band fractionalization effectively splits the electron energy into a distribution of quasiparticle energies. 
The spectral weight of these sub-bands is determined by the phase space of the corresponding operators, implying that it is generally momentum dependent. In strongly correlated systems, this phase space may--to exponential accuracy--vanish, creating voids in parts of the Brillouin zone which change the topology of the excitation spectrum.    
This effect is a prerequisite for Fermi arcs, and spectral topology should therefore be regarded as an essential property of strongly correlated phases.

This work was supported by the Swedish Research Council (VR) through grant 2018-03882. Computations were performed on resources provided by the Swedish National Infrastructure for Computing (SNIC) at the National Supercomputer Centre in Linköping, Sweden.  
\bibliography{biblio.bib}

\clearpage

\section{Appendix I}
\begin{figure*}[!htb]
 \hbox to \linewidth{ \hss
\includegraphics[width=\linewidth]{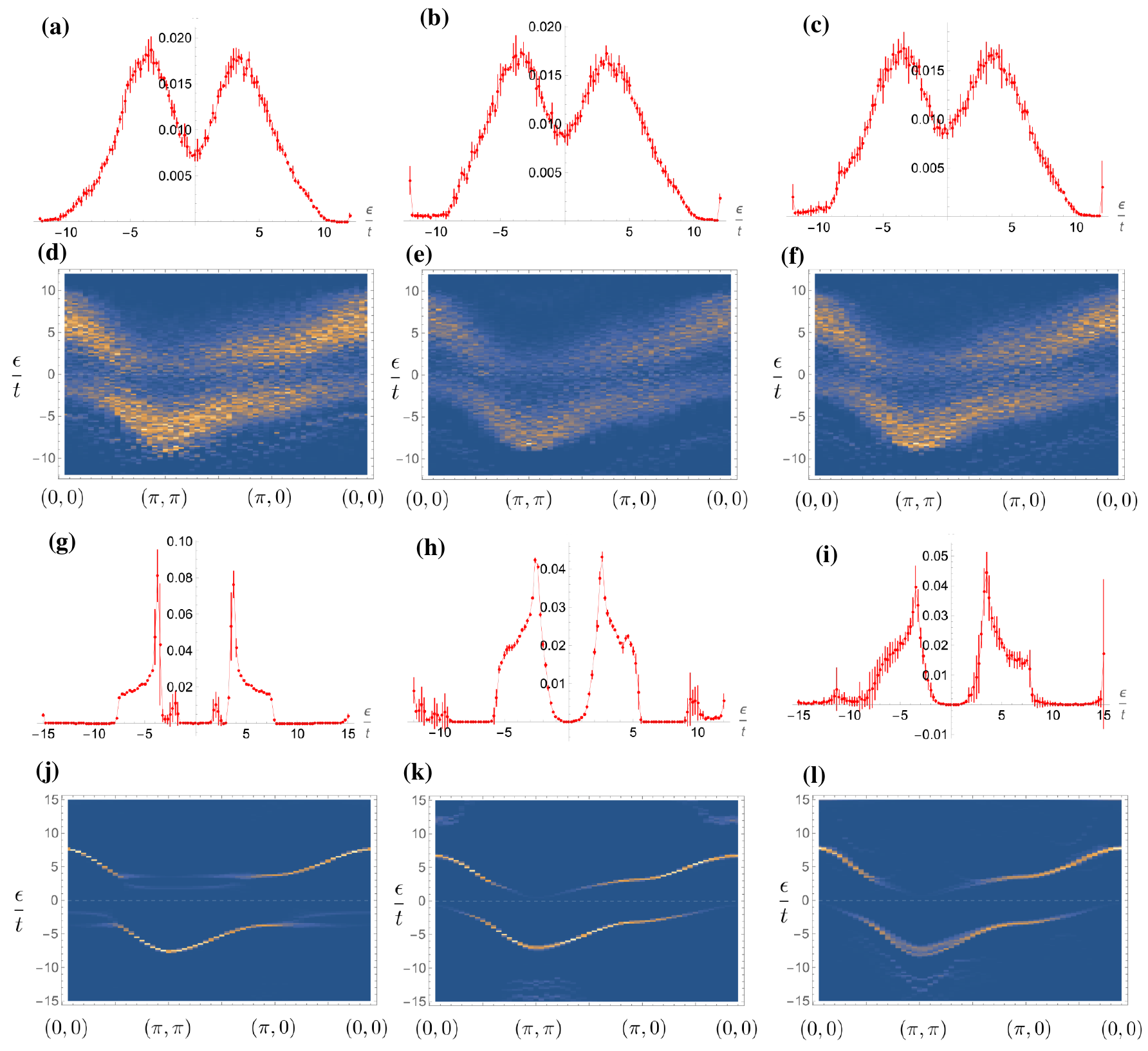}
 \hss}
\caption{
{\bf Convergence of the series at half-filling.} The left column corresponds to an expansion order $O=5$, the center corresponds to $O=6$ and the right corresponds to $O=7$. (a-c) give the dos at a temperature of $t/T=1$, while (d-f) give the corresponding spectra. (g-i) give the dos at a temperature of $t/T=4$, while (j-l) give the corresponding spectra. 
  At the higher temperature, the corrections when changing the expansion order is very small, though a slight shift in dos at the Fermi level can be observed when comparing $O=5$ (a) and $O=6$ (b). 
  At the lower temperature, we do see quantitative difference in dos between orders $5$ (g) and $6$ (h) while the correction at order $7$ (i) is smaller. The small peaks in the dos near the Fermi level in (g) are reflected in a suppressed fractionalized sub-band visible in (j). 
  At orders $6$ and $7$, this fractionalized sub-band vanishes. 
}
\label{A1}
\end{figure*}

To asses how truncation of the series affects the results, we compare the density of states and spectral function for the cases reported in the article at different expansion orders. In Fig. \ref{A1} we show the case of half-filling and temperatures $t/T=1$ and $t/T=4$ for  expansion orders $O=5,6,7$. At the higher temperature, we observe that the dos changes very little, though a small correction at $\epsilon=0$ is visible. The spectrum is qualitatively very similar, and we conclude that the impart of truncation is very small. 

At the lower temperature, we see some changes in the shape of the dos when increasing the order from $5$ to $6$, though the systems consistently remains gapped. The spectra show a weight that does not completely vanish at $O=5$, but is strongly suppressed at higher orders. At $O=7$, we begin to see noise in the spectrum as a result of the computational cost associated with expansions to high order.
For this data set, we can conclude that truncation of the series has a limited quantitative impact, but the it does not affect any of the conclusions derived in the paper.  

\begin{figure*}[!htb]
 \hbox to \linewidth{ \hss
\includegraphics[width=\linewidth]{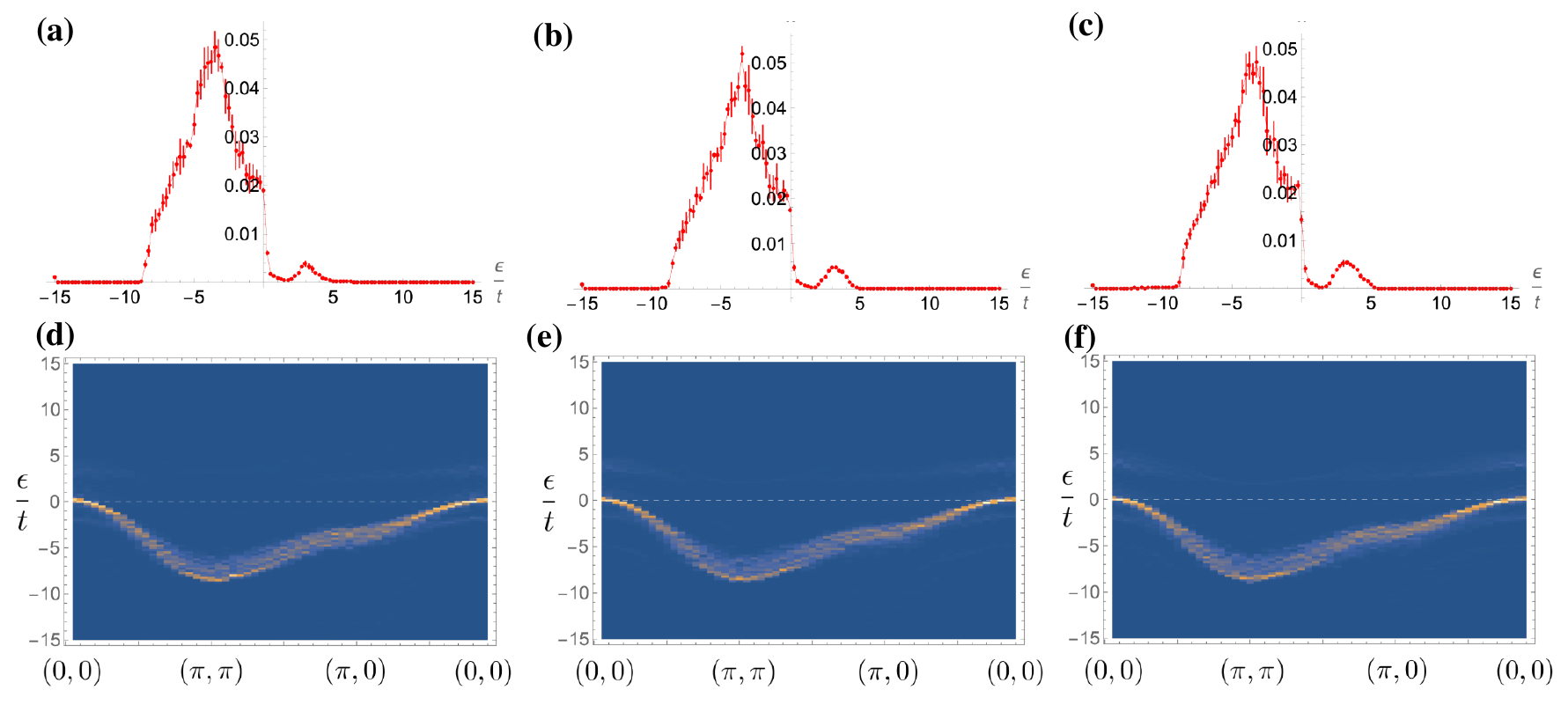}
 \hss}
\caption{
{\bf Convergence of the series in the strongly doped case.}  The density is $\langle \hat{n}\rangle\approx 1.88$ and the temperature is $t/T=4$. The left column (a,d) corresponds to an expansion order $O=5$, the center column to $O=6$ and the right columns to $O=7$.
The dos (a-c) does not change visibly with expansion order, and neither does the spectrum (d-f). We can therefore conclude that the observables have converged. 
}
\label{A2}
\end{figure*}

In Fig. \ref{A2}, we see the dos and spectra for the doped case at expansion orders $O=5,6,7$. In this scenario, truncation of the series has no impact visible to the naked eye, and we can conclude that the result is virtually exact. 
   
In conclusion, we find that the diagrammatic Monte Carlo simulations reported do accurately capture the physics of the attractive Hubbard model. The results are qualitatively not affected by truncation of the series, yet a small quantitative uncertainty remains for one of the data sets. 

\end{document}